\documentclass[aps,12pt,preprintnumbers,nofootinbib,superscriptaddress]{revtex4}
\usepackage{graphicx}
\usepackage{amssymb,amsmath}
\usepackage{slashed}
\usepackage{dcolumn}
\begin{document}

\title{\Large{\bf Note on Weyl versus Conformal Invariance in Field Theory} }

\author{Feng Wu}
\email[Electronic address: ]{fengwu@ncu.edu.cn}
\affiliation{%
Department of Physics, Nanchang University, Nanchang
330031, China}

\date{\today}

\begin{abstract}
It was argued recently that conformal invariance in flat spacetime implies Weyl invariance in a general curved background for unitary theories and possible anomalies in the Weyl variation of scalar operators are identified. We argue that generically unitarity alone is not sufficient for a conformal field theory to be Weyl invariant. Furthermore, we show explicitly that when a unitary conformal field theory couples to gravity in a Weyl invariant way, each primary scalar operator that is either relevant or marginal in the unitary conformal field theory corresponds to a Weyl-covariant operator in the curved background. 
\end{abstract}
\pacs{} 
\maketitle 
\newpage

\date{\today}

\section{Introduction and summary}
Scale and conformal symmetries are essential concepts in quantum field theory. In particular, the renormalization group evolution of a Poincar$\acute{\rm e}$-invariant quantum field theory, being a primary theme of field theory, is controlled by its dynamical behavior under scale transformations. In the study of this subject, the energy-momentum tensor plays a crucial role. In the specific case where a field theory is scale-invariant, it can be shown that the trace of its energy-momentum tensor $T_{\mu}^{\mu} $ must take the form 
\begin{equation}
T_{\mu}^{\mu} = \partial_{\mu} V^{\mu}, \label{virial} 
\end{equation} 
where $V^{\mu}$ is referred to as the ``virial current". If, moreover, the virial current is a total derivative, then the theory is not just a scale-invariant field theory, but it is in fact a conformal field theory. In this case, one can further construct an ``improved" energy-momentum tensor, which is traceless \cite{Callan}.

Although only been proved in two dimensions \cite{Polchinski} and perturbatively in four dimensions \cite{Luty, Dymarsky, Dymarsky2}, it is believed that a Poincar$\acute{\rm e}$-invariant interacting field theory that is scale-invariant but not conformally invariant must be non-unitary. This means that with unitarity, the spacetime symmetry group of a Poincar$\acute{\rm e}$-invariant quantum field theory with scale invariance is enhanced to the conformal group.

In Ref.~\cite{Farnsworth} it is argued that, for unitary theories, conformal invariance in flat spacetime implies local Weyl invariance in a general curved background spacetime. Because of diffeomorphism invariance, a scale transformation of the coordinates and that of the fields in flat spacetime are equivalent to the global Weyl transformations on the metric and fields in a curved spacetime, and hence a quantum field theory with scale invariance in the flat spacetime is globally Weyl invariant when coupled to a general curved background. Thus, it appears that conformal invariance provides a link between global and local Weyl invariance in unitary theories. Early work on this subject includes Refs.~\cite{Polchinski, Iorio,Jackiw,Edery,Brust,Karananas}.

Also, the Weyl transformation of local scalar operators that correspond to primary operators in the flat limit are identified in Ref.~\cite{Farnsworth} and the authors find that there are possible ``anomalous terms" in the transformation formulas that prevent some of these operators from transforming covariantly. They argued that these anomalous terms cannot be eliminated based on the constraints originating from the Abelian nature of the Weyl transformations. 

In this note, with explicit examples provided as demonstration, we show that generally unitarity alone is not sufficient for a conformal field theory to be Weyl invariant. In addition, we show that in the case where a unitary conformal field theory does couple to gravity in a Weyl invariant fashion, each of the relevant and marginal primary scalar operators in the unitary conformal field theory corresponds to a Weyl covariant operator in the curved background. Thus, although the work of this note is highly inspired by Ref.~\cite{Farnsworth}, our analysis has reached different conclusions. 

It is clear that the existence of the local energy-momentum tensor is essential in our analysis. Thus, we have implicitly assumed the existence of the action. Without this assumption, we do not know how to construct a local energy-momentum tensor, not to mention how to couple the theory to gravity. The conclusions of this work may or may not apply to the context where the energy-momentum tensor is not well defined. We leave it for future investigation.

\section{Conformal vs. Weyl}
It is well known that the consequences of symmetries of field theories can be expressed in terms of Ward identities relating Green's functions. For a conformal field theory, the Ward identity for primary operators $O(x)$ under an infinitesimal conformal transformation takes the form
\begin{equation}
\hat{\sigma}(x) \langle T_{\mu}^{\mu}(x) O(x_1)\cdot\cdot\cdot O(x_{n}) \rangle=
\sum_{i}\delta^{(d)}(x-x_{i})\langle O(x_1)\cdot\cdot\cdot( -\Delta \hat{\sigma}(x_{i})O(x_{i})) \cdot\cdot\cdot O(x_{n}) \rangle
\label{WIC},
\end{equation}
where $ \hat{\sigma}(x) ={1\over d} \partial_{\mu} \epsilon^{\mu}(x)$ is the restricted local Weyl rescaling factor with the infinitesimal coordinate change $\epsilon^{\mu}(x)$ given by 
\begin{equation}
\epsilon^{\mu}(x) = a^{\mu} + \omega^{\mu}_{\,\,\nu} x^{\nu}+c x^{\mu} +2(b\cdot x)x^{\mu} -x^2 b^{\mu}
\end{equation}
for translation, Lorentz transformations, scale and special conformal transformations, respectively, in $d$-dimensional flat spacetime. $\Delta$ is the Weyl dimension of the operator $O(x)$. We recall that in flat spacetime the energy-momentum tensor can be generated by the diffeomorphism. That is, under an infinitesimal diffeomorphism
 \begin{equation}
x^{\mu'}=x^{\mu} - \xi^{\mu}(x),
\label{diff}
\end{equation}
the action transforms as
 \begin{equation}
\delta S= {1\over 2} \int d^{d} x (\partial_{\mu}\xi_{\nu}+\partial_{\nu}\xi_{\mu}) T^{\mu\nu}. \label{EM1}
\end{equation}

On the other hand, when the theory is coupled to a general curved metric $g_{\mu\nu}$, the energy-momentum tensor can also be determined by the response of the action to a local variation of the metric. Explicitly, under the variation
\begin{equation}
g_{\mu\nu}\rightarrow g_{\mu\nu} +\delta g_{\mu\nu},
\end{equation}
we have
\begin{equation}
\delta S= -{1\over 2} \int d^{d} x \sqrt{\vert g\vert}\delta g_{\mu\nu} T^{\mu\nu}. \label{EM2}
\end{equation}
This is consistent with the expression of Eq.~(\ref{EM1}) in flat space by general covariance.

For a Weyl-invariant theory, it is straightforward to show that the response of the $n$-point correlator for $O(x)$ to an infinitesimal Weyl transformation $\delta g_{\mu\nu}(x)=2\sigma(x) g_{\mu\nu}(x)$ in odd dimensions contains only contact terms:
\begin{equation}
\sigma(x) \langle T_{\mu}^{\mu}(x) O(x_1)\cdot\cdot\cdot O(x_{n}) \rangle=
\sum_{i}\delta^{(d)}(x-x_{i})\langle O(x_1)\cdot\cdot\cdot \delta_{\sigma}O(x_{i}) \cdot\cdot\cdot O(x_{n}) \rangle
\label{WIW1},
\end{equation}  
where $ \delta_{\sigma}O(x_{i})$ is the variation of the operator $O(x_{i})$ under the infinitesimal Weyl transformation. Meanwhile, due to the Weyl anomaly \cite{Capper, Deser}, the Weyl Ward identity in even dimensions is modified to take the form
\begin{equation}
\sigma(x) \langle T_{\mu}^{\mu}(x) O(x_1)\cdot\cdot\cdot O(x_{n}) \rangle=
\sum_{i}\delta^{(d)}(x-x_{i})\langle O(x_1)\cdot\cdot\cdot \delta_{\sigma}O(x_{i}) \cdot\cdot\cdot O(x_{n}) \rangle + \sigma(x) \langle \mathcal{A}(x) O(x_1)\cdot\cdot\cdot O(x_{n}) \rangle
\label{WIW2},
\end{equation}  
where the local function $\mathcal{A}(x)$ stands for the Weyl anomaly terms.

In Ref.~\cite{Farnsworth}, it is argued that conformal invariance in flat spacetime implies Weyl invariance in a general curved background for unitary theories. Showing that a conformal field theory in flat spacetime is Weyl invariant in a curved background metric is equivalent to showing that Eq.~(\ref{WIC}) implies Eqs.~(\ref{WIW1}) and~(\ref{WIW2}) in odd and even dimensions, respectively. The argument begins with the statement that because the ``improved" energy-momentum tensor $T^{\mu\nu}$ vanishes for a unitary conformal field theory in flat spacetime, the theory coupling to gravity in curved spacetime would have $T^{\mu\nu}$ proportional to at least one power of the Riemann curvature tensor $R$. Constraints from unitarity and commutativity of Weyl transformations are then used to eliminate all possible contributions to $T^{\mu\nu}$. If correct, the above argument would imply that a unitary non-Weyl-invariant theory in a curved background cannot be a conformal field theory in the flat space limit.

Now, we discuss the potential loophole in the above argument. First, we should note that if one uses Eq.~(\ref{EM2}) to calculate the energy-momentum tensor of a theory and find a traceless one, it means that the theory being considered has already coupled to gravity in a Weyl-invariant way. In this case, it is not meaningful to use the argument of Ref.~\cite{Farnsworth} to show that this theory is Weyl invariant, since it would seem that one is attempting to shut a box that is already closed. 

The better question is: given a unitary conformal field theory whose energy-momentum tensor is generated by the diffeomorphism in flat space, does conformal invariance along with unitarity implies Weyl invariance in curved space? 

Given a conformal field theory in flat space, there is no unique way to couple it to gravity. Indeed, this ambiguity is the origin of the improvement of the energy-momentum tensor. Thus, $T_{\mu}^{\mu}=0$ in flat space does not guarantee $T_{\mu}^{\mu}=\mathrm{O}(R)$ in curved space. For example, there could be terms in the Lagrangian of a conformal field theory that generate nonvanishing contributions to the energy-momentum tensor under the diffeomorphism in flat space, but whose Weyl variations in a curved space are surface terms. 

To be more explicit, consider the action of a free massless scalar $\phi$ given by 
\begin{equation}
 S= \int d^{d} x \left({1\over 2} \partial_{\mu}\phi \partial^{\mu} \phi - {d-2 \over 4(d-1)} \partial^{2} \phi^2\right). \label{ex1}
\end{equation}
This unitary theory is conformally invariant, since the variation of the second term by a diffeomorphism, Eq.~(\ref{diff}), generates the ``improved" contribution to the traceless energy-momentum tensor \cite{Hill}. Note that although the second term in Eq.~(\ref{ex1}) is a surface term, and thus not affecting the equation of motion, it varies under the diffeomorphism and produces a nonzero contribution to the energy-momentum tensor. 

However, when minimally coupled to a background metric, it is straightforward to show that the action 
\begin{equation}
 S= \int d^{d} x \sqrt{\vert g\vert} \left({1\over 2} g_{\mu\nu}\partial^{\mu}\phi \partial^{\nu} \phi - {d-2 \over 4(d-1)} \nabla_{\mu}\partial^{\mu} \phi^2\right)
\end{equation}
is not Weyl invariant unless $d=2$. In fact, under an infinitesimal  Weyl variation $\delta g_{\mu\nu}=2\sigma g_{\mu\nu}$ and $\delta \phi = -{d-2\over 2}\sigma \phi$, the action transforms as
\begin{equation}
\delta_{\sigma} S={d-2\over 4} \int d^{d} x \sqrt{\vert g\vert} (\square \sigma) \phi^{2}.
\end{equation} 
This is an example showing that a non-Weyl invariant theory can reduce to a unitary conformal field theory in the flat space limit. In other words, we can have a unitary conformal field theory that couples to gravity in a non-Weyl-invariant way.

Certainly, there exist other possibilities such that a unitary conformal field theory cannot couple to gravity in a Weyl-invariant way. For example, a theory can fail to be Weyl invariant in the curved background because of the specific symmetry that prevents one from constructing the would-be Weyl-invariant Lagrangian. A free massless scalar with the shift symmetry $\phi\rightarrow \phi +c$ is such an example \cite{Dymarsky, Nakayama}. In this case, the well-known ``improved" coupling term ${d-2\over 8(d-1)} R\phi^2$ with $R$ being the Ricci scalar is not allowed to be included in the action by symmetry, and thus this theory is not Weyl invariant in curved spacetime unless $d=2$. Therefore, unitarity alone is not sufficient for a conformal field theory to be Weyl invariant.

\section{Contact terms}
Having shown that a unitary conformal field theory might not couple to gravity in a Weyl-invariant way, we will now concentrate our attention to the specific situation of interest where a conformal field theory does couple to gravity in a Weyl-invariant way and consider contact terms in Eqs.~(\ref{WIW1}) and~(\ref{WIW2}).
   
As described in Ref.~\cite{Farnsworth}, one must have $ \delta_{\sigma}O\rightarrow -\Delta \hat{\sigma} O $ in the flat limit and $ \sigma\rightarrow \hat{\sigma}$, with $\hat{\sigma}$ given below Eq.~(\ref{WIC}). In the special case where $O$ does not contain the metric tensor $g_{\mu\nu}$, $\delta_{\sigma} O$ must transform covariantly, that is, $ \delta_{\sigma} O = -\Delta \sigma O$. The reason that the Weyl variation of the scalar operator $O$ does not contain terms involving the derivatives of $\sigma$ is simply because, without the metric tensor, no scalar operator can be formed out of derivatives of $\sigma$.

Now, let us consider the general case where $O$ consists of matter fields, the metric tensor and their derivatives. As already mentioned above, scale transformations in flat spacetime are equivalent to global Weyl transformations in the curved background. Thus, when $\sigma=c$ with $c$ being a constant, we shall have
\begin{equation}
\delta_{\sigma=c} O= -\Delta c O,
\end{equation}    
from which it follows that under a general Weyl transformation, the operator $O$ transforms either covariantly or as 
\begin{equation}
\delta_{\sigma} O= -\Delta \sigma O + \mathsf{O}(\partial \sigma) \label{variationO}.
\end{equation}  
Note that the first term in the variation Eq.~(\ref{variationO}) is the only permitted term that is proportional to $\sigma$. Terms that violate Weyl covariance are at least of order $\partial \sigma$. Terms such as $\sigma R^2 U$ or $\sigma W^{\mu\nu\alpha\beta}W_{\mu\nu\alpha\beta} U$ (where the shorthand notation $R$ stands for the curvature tensor, the Ricci tensor and the Ricci scalar, $W_{\mu\nu\alpha\beta}$ is the Weyl tensor, and $U$ is a scalar operator with Weyl dimension $\Delta-4$), referred to as the ``anomalous terms" in \cite{Farnsworth}, are not allowed unless the operator $O$ is itself proportional to $ R^2 U$ or $ W^{\mu\nu\alpha\beta}W_{\mu\nu\alpha\beta} U$.  

Then, requiring the Wess-Zumino consistency condition \cite{Wess} for the Weyl variation, that is, $[ \delta_{\sigma_{1}},  \delta_{\sigma_{2}}]O=0$, the most general Weyl variation of $O$ allowed by symmetries and unitarity constraints on the dimensions of operators can be identified. Since we do not know of any example of an interacting conformal theory with spacetime dimension $d>6$, we will restrict our attention to spacetime dimension $d \leq 6$. The calculations are straightforward but not very illuminating. The results for relevant and marginal scalar operators in $d \leq 6$ are presented as follows.  

For $\Delta \geq {d+2\over 2}$ and $\Delta \neq 2n$, $n=1,2,3$, we have
\begin{equation}
\delta_{\sigma} O= -\Delta \sigma O + A \Box \sigma \label{variationOg},
\end{equation} 
where $A$ is a Weyl covariant scalar with Weyl dimension $\Delta_{A}=\Delta-2$. As shown in \cite{Farnsworth}, the new operator $O'$ defined as
\begin{equation}
 O'=O+{1\over 2(d-1)} RA
\end{equation} 
 transforms covariantly as $\delta_{\sigma} O'= -\Delta \sigma O'$.

Operators with $\Delta=2n$ are special. For $\Delta=2$, the variation reads
 \begin{equation}
\delta_{\sigma} O_{2}= -2 \sigma O_{2} + c_{1} \Box \sigma. \label{variationO2}
\end{equation} 
For $\Delta=4$, we have
 \begin{equation}
\delta_{\sigma} O_{4}= -4 \sigma O_{4} +B\Box \sigma+c_{2} R \Box \sigma \,\,\,\,\,\,\, {\rm{in}} \,\,\,d=4,5, \label{variationO41}
\end{equation} 
with the Weyl dimension 2 operator $B$ transforming according to $ \delta_{\sigma} B= -2 \sigma B +c_{1}'\Box \sigma$, whereas
\begin{equation}
\delta_{\sigma} O_{4}= -4 \sigma O_{4} +B\Box \sigma+c_{2} R \Box \sigma +c_{3} \Box^2 \sigma \,\,\,\,\,\,\, {\rm{in}} \,\,\,d=6. \label{variationO42}
\end{equation}
Note that the term involving $\Box^2 \sigma$ is allowed only in $d=6$. This is due to the fact that under the Weyl variation,
 \begin{equation}
\delta_{\sigma_{2}} \Box^2 \sigma_{1}= -4 \sigma_{2}\Box^2 \sigma_{1}+(d-6)g^{\mu\nu}\nabla_{\nu}(\Box \sigma_{1})\nabla_{\mu}\sigma_{2}-2\Box\sigma_{1}\Box\sigma_{2}+(d-2)\Box( g^{\mu\nu}\nabla_{\mu}\sigma_{1}\nabla_{\nu}\sigma_{2}  ).
\end{equation} 
Thus, if the variation $\delta_{\sigma} O_{4}$ contains the term involving $\Box^2 \sigma$, the commutativity of Weyl transformations cannot be satisfied unless $d=6$.

Finally, for $\Delta=d=6$, we have
\begin{equation}
\delta_{\sigma} O_{6}= -6 \sigma O_{6} +A'\Box \sigma+B' \Box^2 \sigma + B'' R \Box \sigma +c_{4} R^2 \Box \sigma , \label{variationO6}
\end{equation}
where the Weyl variations of the operators $A'$, $B'$ and $B''$ are given, respectively, by 
\begin{equation}
\delta_{\sigma} A'= -4 \sigma A' +B'''\Box \sigma+c_{5} R \Box \sigma 
\end{equation}
with $ \delta_{\sigma} B'''= -2 \sigma B''' +c_{1}'''\Box \sigma$,
\begin{equation}
\delta_{\sigma} B'= -2 \sigma B',
\end{equation}
and
\begin{equation}
\delta_{\sigma} B''= -2 \sigma B'' +c_{1}''\Box \sigma. 
\end{equation}

Now, let us introduce the operators
\begin{equation}
 O_{2}'\equiv O_{2}+{c_{1}\over 2(d-1)} R,
\end{equation} 
\begin{equation}
 O_{4}'\equiv O_{4}+{1\over 2(d-1)} RB+{1\over 4(d-1)}(c_{2}+{c_{1}' \over 2(d-1)})R^2 \,\,\,\,\,\,\, {\rm{in}} \,\,\,d=4,5,
\end{equation} 
\begin{equation}
 O_{4}''\equiv O_{4}+{1\over 10} RB+{1\over 20}(c_{2}+{c_{1}' \over 10}-{c_{3}\over 5})R^2 +{c_{3} \over 10}\Box R \,\,\,\,\,\,\, {\rm{in}} \,\,\,d=6,
\end{equation} 
and
\begin{equation}
 O_{6}'\equiv O_{6}+{1\over 10}A' R+{1\over 20}\left(-{1\over 5} B'+ B'' +{1\over 10} B'''\right)R^2 +  {1 \over 10}B'\Box R+{1\over 30}\left(c_{4}+{c_{5}\over 10}+{c_{1}''\over 20}+{c_{1}'''\over 200}\right)R^3 ,
\end{equation} 
it is straightforward to show that operators $O_{2}'$, $O_{4}'$, $O_{4}''$ and $O_{6}'$ all transform covariantly under an infinitesimal Weyl transformation.

With these results, we conclude that when a conformal field theory in $d\leq 6$ is coupled to a general curved background metric $g_{\mu\nu}$ in a Weyl-invariant way, every primary scalar operator $O(x)$ that is either relevant or marginal corresponds to a Weyl-covariant operator $O'(x)$ such that $O'(x)\rightarrow O(x)$ in the flat limit, and the operators $O'(x)$ obey the infinitesimal form of the Ward identities for Weyl invariance given by
\begin{equation}
\sigma(x) \langle T_{\mu}^{\mu}(x) O'(x_1)\cdot\cdot\cdot O'(x_{n}) \rangle=
\sum_{i}\delta^{(d)}(x-x_{i})\langle O'(x_1)\cdot\cdot\cdot (-\Delta \sigma(x_{i}) O'(x_{i})) \cdot\cdot\cdot O'(x_{n}) \rangle.
\end{equation}      
and
\begin{eqnarray}
\sigma(x) \langle T_{\mu}^{\mu}(x) O'(x_1)\cdot\cdot\cdot O'(x_{n}) \rangle&=
&\sum_{i}\delta^{(d)}(x-x_{i})\langle O'(x_1)\cdot\cdot\cdot (-\Delta \sigma(x_{i}) O'(x_{i})) \cdot\cdot\cdot O'(x_{n}) \rangle \nonumber \\
&+& \sigma(x) \langle \mathcal{A}(x) O'(x_1)\cdot\cdot\cdot O'(x_{n}) \rangle
\end{eqnarray}  
in odd and even dimensions, respectively.

\begin{acknowledgments}
The author is grateful to S. Deser and M. Luty for correspondence. This research was supported in part by the National Nature Science Foundation of China under Grant Nos. 11665016 and 11565019.
\end{acknowledgments}



\end{document}